\documentclass{revtex4-2}
\usepackage[utf8]{inputenc}
\usepackage{amsbsy}
\usepackage{amsopn}
\usepackage{amstext}
\usepackage{amsmath,amsthm,amsfonts,amssymb}
\usepackage[mathcal]{eucal}
\usepackage{mathrsfs}
\usepackage[english]{babel}
\usepackage{color}
\usepackage{esint}
\usepackage{graphicx}
\usepackage{float}
\usepackage{units}
\usepackage{textcomp}

\DeclareGraphicsExtensions{.png,PNG,.pdf,.PDF}
\usepackage{hyperref}
\usepackage{slashed}
\newcommand{\ie}{\begin{equation}}
\newcommand{\fe}{\end{equation}}
\newcommand{\se}{\begin{eqnarray}}
\newcommand{\ff}{\end{eqnarray}}
\begin{document}

\title{Thermodynamic properties of the noncommutative quantum Hall effect with anomalous magnetic moment}
\author{R. R. S. Oliveira}
\email{rubensrso@fisica.ufc.br}
\affiliation{Universidade Federal do Cear\'a (UFC), Departamento de F\'isica,\\ Campus do Pici, Fortaleza - CE, C.P. 6030, 60455-760 - Brazil.}


\author{R. R. Landim}
\email{renan@fisica.ufc.br}
\affiliation{Universidade Federal do Cear\'a (UFC), Departamento de F\'isica,\\ Campus do Pici, Fortaleza - CE, C.P. 6030, 60455-760 - Brazil.}


\date{\today}

\begin{abstract}
In this paper, we study the thermodynamic properties of the noncommutative quantum Hall effect (NCQHE) with anomalous magnetic moment (AMM) for both relativistic and nonrelativistic cases in the high temperatures regime. Thus, we use the canonical ensemble for a set of $N$-particles in contact with a thermal bath. Next, we explicitly determine the thermodynamic properties of our interest, namely: the Helmholtz free energy, the entropy, the mean energy, and the heat capacity. In order to perform the calculations, we work with the \textit{Euler-MacLaurin} formula to construct the partition function of the system. In that way, we plotted the graphs of thermodynamic properties as a function of temperature for six different values of the magnetic field and of the NC parameters. As a result, we note that the Helmholtz free energy decreases with the temperature, increases with the NC parameters, and can decrease or increase for certain values of the magnetic field, white that the entropy increases with the temperature, decreases with the NC parameters, and can decrease or increase for certain values of the magnetic field. Besides, the mean energy increases linearly with the temperature and its values for the relativistic case are twice of the nonrelativistic case, consequently, the heat capacity for the relativistic case is twice of the nonrelativistic case, where both are constants, and therefore, satisfying the so-called \textit{Dulong-Petit} law. Finally, we also verify that there is no influence of the AMM on the thermodynamic properties of the system.
\end{abstract}


\maketitle

\section{Introduction}

In 1980, K. von Klitzing discovered the quantized version of the classical Hall effect (CHE), or simply Hall effect (HE), in what became known as the quantum Hall effect (QHE) \cite{K1,K2,Cage,Yoshioka}. For this discovery, von Klitzing was awarded the 1985 Nobel Prize in Physics. Unlike the CHE, the QHE is only observed when a two-dimensional electron gas (2DEG) is subjected to low temperatures and strong magnetic fields, consequently, the energy spectrum (Landau levels), the electrical conductivity (Hall conductivity), and the electrical resistivity (Hall resistivity) are quantized quantities (discrete or discontinuous). In fact, for weak fields and high temperatures (room temperature), the QHE reduces to the CHE. However, in 1988 a model for a QHE without Landau levels was proposed by F. D. M. Haldane (2016 Nobel Prize in Physics) \cite{Haldane,Hang}, known as the quantum anomalous Hall effect, and in 2007 the (integer) QHE was reported in graphene at temperatures as high as room temperature (graphene is an excellent host of a 2DEG) \cite{Novoselov,Wu}. Also, the QHE has already been studied in nonrelativistic and relativistic quantum mechanics (low and high energy regime) \cite{Frohlich,Ishikawa,Thouless,Filgueiras,Haldane,Schakel,Lamata}.

In 1947, H. S. Snyder published two papers on quantized spacetimes, thus introducing the concept of noncommutative (NC) space as we know it today \cite{Snyder1,Snyder2}. For Snider, although the Minkowski spacetime is a continuum, this assumption is not required by Lorentz invariance, and therefore, a model of a Lorentz invariant discrete spacetime inspired by quantum mechanics was perfectly possible (in theory). Later, an even more general concept emerged, namely, the NC phase space, which obeys a NC algebra \cite{Szabo,Douglas,Gomis,Seiberg}. In essence, a NC phase space is based on the assumption that the position and momentum operators are NC variables that must satisfy $[x^{NC}_i,x^{NC}_j]=i\theta_{ij}$ and $[p^{NC}_i,p^{NC}_j]=i\eta_{ij}$ (space-like NC), where $\theta_{ij}=\theta\epsilon_{ij}$ and $\eta_{ij}=\eta\epsilon_{ij}$, being $\theta>0$ and $\eta>0$ the position and momentum NC parameters with dimensions of length-squared and momentum-squared, respectively \cite{Berto,Bastos}. From a phenomenological point of view, supposed signatures of NC were investigated in the decay of kaons, photon-neutrino interaction, etc \cite{Hinchliffe,Melic,Schupp,Abel,Pikovski}. Furthermore, NC (phase) spaces can be applied in quantum chromodynamics (QCD) \cite{Carlson}, quantum electrodynamics (QED) \cite{Riad}, black holes \cite{Nicolini}, standard model \cite{Melic}, quantum cosmology \cite{Garcia}, graphene \cite{Bastos,Santos}, etc.

In QED, the anomalous magnetic moment (AMM) of a Dirac fermion is a contribution of effects of the quantization of the electromagnetic field or vacuum polarization (expressed by Feynman diagrams with loops) to the total magnetic dipole moment (MDM) of that fermion \cite{Greiner1,Greiner2,Geiger}. In particular, the AMM was first found by J. S. Schwinge in 1948 (1965 Nobel Prize in Physics with Feynman and Tomonaga) and is represented by: $a=a^{QED}=\frac{\alpha}{2\pi}\approx 0.0011614$ (one-loop result), where $\alpha\approx\frac{1}{137}$ is the famous fine-structure constant \cite{Schwinger}. Currently, the analytical (theoretical) value of the AMM calculated up to the order $\alpha^5$ is given by: $a_e=0.00115965218178(77)$ \cite{Aoyama}, while the experimental value is given by: $a_e=0.00115965218073(28)$ \cite{Hanneke}, and therefore, the QED result agrees with the experimental result of more than 10 significant figures, making the AMM (of the electron) the most accurately verified prediction in the history of physics. On the other hand, in addition to electrons, practically all Dirac fermions have a characteristic AMM \cite{Greiner1,Greiner2,Geiger,Jegerlehner,Abi,Acciarri}. Furthermore, fermions with AMM are the basis for modeling the Aharonov-Casher effect \cite{Aharonov,Oliveira1,Oliveira2}, and the Dirac oscillator \cite{Moshinsky,Martinez,Oliveira3,Oliveira4}. Recently, the noncommutative quantum Hall effect (NCQHE) with AMM has been studied in three different relativistic scenarios \cite{Oliveira5}.

The study of the physical properties of materials, focusing primarily on their thermodynamic (thermal) properties, is of great interest in physics, such as in condensed matter physics, solid-state physics, and materials science, respectively \cite{Gaskell,DeHoff,Tester,Dolling,Muhlschlegel,Coleman,Eckert,Foiles,Anthony,Wang,Grimvall}. Indeed, the practical relevance of such properties can be found in Refs. \cite{Balandin,Mounet,Shahil,Pop,Alofi,Che,Ruoff,Philip}, where the thermodynamic properties of graphene, diamond, graphite, carbon nanotubes, nanostructured carbon materials, and nanofluids were investigated. In addition, the study of such properties is also of great relevance from a theoretical point of view, i.e., for low and high-energy quantum systems. For example, in relativistic quantum mechanics, we can investigate the thermodynamic properties of the Dirac oscillator \cite{Pacheco1,Pacheco2,Oliveira6}, neutral Dirac fermions with AMM \cite{Oliveira7}, Aharonov-Bohm quantum rings \cite{Oliveira8}, two-dimensional relativistic gas \cite{Montakhab}, Klein-Gordon oscillator \cite{Hassanabadi}, etc. Already in nonrelativistic quantum mechanics, we can investigate the thermodynamic properties of quantum pseudodots and dots \cite{Groote,Donfack}, quantum wires \cite{Najafi}, quantum wells \cite{Olendski}, NC graphene \cite{Santos}, diatomic molecules \cite{Edet}, etc.

This paper has as its goal to study the thermodynamic properties of the NCQHE with AMM for both relativistic and nonrelativistic cases in the high temperatures regime. To achieve this, we use the canonical ensemble as a statistical ensemble to accommodate a set of noninteracting $N$-particles in contact with a thermal bath (heat reservoir) at a constant temperature $T$ (thermodynamic equilibrium temperature). In that way, we have a closed system characterized by a constant number of particles ($N=cte$) confined in a volume also constant ($V=cte$) in which only heat and work (non-PV work) can be exchanged between the system ($N$-particles) and the thermal bath (both isolated from the rest of the universe) \cite{Greiner,Pathria}. Next, we explicitly determine the (macroscopic) thermodynamic properties of our interest, namely: the Helmholtz free energy, the entropy, the mean (internal) energy, and the heat capacity. Also, in order to perform the calculations of these properties, we use the \textit{Euler-MacLaurin} formula to construct the (canonical) partition function of the system.

This paper is organized as follows. In Section \ref{sec2}, we explicitly determine the thermodynamic properties of the NCQHE with AMM for both relativistic and nonrelativistic cases in the high temperatures regime. In Section \ref{sec3}, we present the results and discussions, where we analyze in detail through 2D graphs the behavior of thermodynamic properties as a function of temperature for six different values of the magnetic field $B$ and of the NC parameters $\theta$ and $\eta$. In Section \ref{conclusion}, we finish our work with the conclusion.

\section{Thermodynamic properties of the system \label{sec2}}

In this section, we calculate the relativistic and nonrelativistic thermodynamic properties of a $N$-particles system in contact with a thermal bath at temperature $T$. According to statistical thermodynamics, these properties are the Helmholtz free energy $F(T,V,N)$, the entropy $S(T,V,N)$, the mean energy $U(T,V,N)$, and the heat capacity $C_V(T,V,N)$, where such properties are calculated from the partition function and are presented below.

\subsection{The relativistic case\label{subsec1}}

As a starting point, we have the following relativistic energy spectrum of the NCQHE with AMM in the Minkowski spacetime (in polar coordinates with $\hbar=c=1$) \cite{Oliveira5}
\begin{equation}\label{spectrum}
E^{\kappa}_{n,m_j,s}=E_m+\kappa\sqrt{m_0^2+2m_0\tau\lambda\omega_c N},
\end{equation}
with
\begin{equation}\label{M19}
E_m=\mu_mB>0, \ \	N=N_{eff}=\left(n+\frac{1}{2}+\frac{\big|m_j-\frac{s}{2}\big|+\left(m_j+\frac{s}{2}\right)}{2}\right),
\end{equation}
and
\begin{equation}\label{taulambda}
\tau\lambda=\frac{1}{4m_0\omega_c}(4-m_0\omega_c\theta)(m_0\omega_c-\eta)>0,
\end{equation}
where $E_m$ is the magnetic (potential) energy of the fermion with a MDM given by $\mu_m=a\mu_B$, where $a$ is its AMM and $\mu_B=\frac{e}{2m_0}$ is the Bohr magnetic (MDM quantum), with $B>0$ being the strength of the uniform external magnetic field (or simply a magnetic field), and $e$ is the electric of the fermion with a rest mass $m_0$, $N$ is an effective quantum number, being $n=0,1,2,\ldots$ the (radial) quantum number and $m_j=\pm\frac{1}{2},\pm\frac{3}{2},\ldots$ the total magnetic quantum number, $s=\pm 1$ is the spin parameter (describes the spin ``up'' or ``down''), $\theta$ and $\eta$ are the position and momentum NC parameters, $\omega_c=\frac{eB}{m_0}$ is the cyclotron frequency (an angular velocity), and $\kappa=\pm 1$ is a parameter that describes the positive or negative energy states (particle or antiparticle), respectively. So, from a physical point of view, this spectrum has a characteristic that allows us to calculate the thermodynamic properties of the system (for particles), which is the fact of having a finite degeneracy (finite degenerate states) for $m_j>0$ (regardless of the spin chosen). Therefore, considering that $\kappa=s=+1$, we have the following finitely degenerate spectrum for the NCQHE \cite{Oliveira5}
\begin{equation}\label{spectrum2}
E_{k}=E_m+m_0\sqrt{\Bar{A}+Ak}>0,
\end{equation}
where $k=n+m_l\geq 0$ is a new quantum number, being $m_l\geq 0$ the orbital magnetic quantum number (arises from $m_j=m_l+s/2$), and the parameters $\Bar{A}$ and $A$ are given by: $\Bar{A}=1+A$ and $A=\frac{2\tau\lambda\omega_c}{m_0}$.

Now, let's focus our attention on the fundamental object of the statistical mechanics (for the canonical ensemble), i.e., the partition function $Z$, which is defined as the sum of all possible macroscopic quantum states of the system \cite{Greiner,Pathria}. Explicitly, the one-particle partition function ($N=1$) is given by the following expression \cite{Pacheco2,Greiner,Pathria,Oliveira4}
\begin{equation}\label{partition}
Z(T,V,1)=\sum_{k=0}^\infty\Omega(E_k)e^{-\beta E_k},
\end{equation}
where $\beta=\frac{1}{k_B T}$ is the Boltzmann factor, with $k_B$ being the Boltzmann constant, and the quantity $\Omega(E_k)$ is the degree of degeneracy (or simply the degeneracy) for the energy level $E_k$ (number of microstates of the system with energy $E_k$). To determine $\Omega(E_k)$, it is necessary to take into account that for each quantum level (state) described for a specific pair ($n,m_l$), there are $2m_l+1$ different degenerate states \cite{Pacheco2,Oliveira4,Oliveira5}. In this way, the total degree of degeneracy (total number of microstates) is given by the following equation
\begin{equation}\label{degeneracy}
\Omega(E_k)=\sum_{m_l=0}^k(2m_l+1)=(k+1)^2.
\end{equation}

Therefore, the partition function \eqref{partition} becomes
\begin{equation}\label{partition2}
Z(T,V,1)=\sum_{k=0}^\infty (k+1)^2e^{-[\bar{\beta}+\tilde{\beta}\sqrt{\Bar{A}+Ak}]},
\end{equation}
where $\bar{\beta}=\beta E_m$ and $\tilde{\beta}=\beta m_0$. 

Before proceeding, it is advisable to analyze the convergence of the partition function \cite{Pacheco1,Pacheco2,Oliveira4}. So, the function $f(x)=(x+1)^2e^{-[\bar{\beta}+\Tilde{\beta}\sqrt{\Bar{A}+Ax}]}$ is a monotonically decreasing function if the following associated integral
\begin{eqnarray}\label{integral}
I(\bar{\beta},\Tilde{\beta})&=&\int_{0}^{\infty}f(x)dx=\left[\frac{240}{A^3\Tilde{\beta}^6}+\frac{240\sqrt{\Bar{A}}}{A^3\Tilde{\beta}^5}+\frac{(24A+96\Bar{A})}{A^3\Tilde{\beta}^4}\right]e^{-[\bar{\beta}+\Tilde{\beta}\sqrt{\Bar{A}}]} \nonumber \\
&& +\left[\frac{(192A+128\Bar{A})\sqrt{\Bar{A}}}{A^3\Tilde{\beta}^3}+\frac{(8A+2\Bar{A})}{A^2\Tilde{\beta}^2}+\frac{2\sqrt{\bar{A}}}{A\Tilde{\beta}}\right]e^{-[\bar{\beta}+\Tilde{\beta}\sqrt{\Bar{A}}]},
\end{eqnarray}
is finite (convergent). From the theorems of convergent series, this result is finite and, therefore, the partition function \eqref{partition2} is also finite.

 However, although the partition function \eqref{partition2} is a convergent function, it cannot be calculated exactly in a closed form \cite{Greiner,Pathria}. On the other hand, for high temperatures ($T\to\infty$ or $\beta\ll 1$) we can get a good approximation \cite{Greiner,Pathria}. So, a systematic expansion of \eqref{partition2} for large $T$ is possible with the use of the \textit{Euler-MacLaurin} (sum) formula, in which the objective is to calculate the integrals numerically. Explicitly, the \textit{Euler-MacLaurin} formula is given by \cite{Greiner,Pathria}
\begin{equation}\label{partition3}
Z(T,V,1)=\sum_{k=0}^\infty f(k)=\frac{1}{2}f(0)+\int_{0}^{\infty}f(x) dx-\sum_{p=1}^{\infty}\frac{1}{(2p)!}B_{2p}f^{(2p-1)}(0),
\end{equation}
or
\begin{equation}\label{partition4}
 Z(T,V,1)=\sum_{k=0}^{\infty}f(k)=\frac{1}{2}f(0)+\int_{0}^{\infty}f(x) dx-\frac{1}{12}f'(0)+\frac{1}{720}f'''(0)-\ldots+,
\end{equation}
where $B_{2p}$ are the Bernoulli numbers.

Using the information above, the partition function \eqref{partition4} is rewritten as follows
\begin{eqnarray}\label{partition5}
Z(T,V,1)&=&\left[\frac{1}{2}+\frac{240}{A^3\Tilde{\beta}^6}+\frac{240\sqrt{\Bar{A}}}{A^3\Tilde{\beta}^5}+\frac{(24A+96\Bar{A})}{A^3\Tilde{\beta}^4}+\frac{(192A+128\Bar{A})\sqrt{\Bar{A}}}{A^3\Tilde{\beta}^3}\right]e^{-[\bar{\beta}+\Tilde{\beta}\sqrt{\bar{A}}]} \nonumber \\
&& +\left[\frac{(8A+2\Bar{A})}{A^2\Tilde{\beta}^2}+\frac{2\sqrt{\bar{A}}}{A\Tilde{\beta}}-\frac{(4\sqrt{\bar{A}}-A\Tilde{\beta})}{24\sqrt{\bar{A}}}-\frac{A\Tilde{\beta}(24\Bar{A}^2-12\Bar{A}A+3A^2)}{5.760\bar{A}^{5/2}}\right]e^{-[\bar{\beta}+\Tilde{\beta}\sqrt{\bar{A}}]} \nonumber \\
&& +\left[\frac{\sqrt{\Bar{A}}\Tilde{\beta}(12\Bar{A}^2A-3A^2\Bar{A})-\Bar{A}A^3\Tilde{\beta}^3}{5.760\bar{A}^{5/2}}\right]e^{-[\bar{\beta}+\Tilde{\beta}\sqrt{\bar{A}}]}+O(\Tilde{\beta}^4). \nonumber\\
\end{eqnarray}

Therefore, in the high temperatures regime, the partition function \eqref{partition5} takes the form
\begin{equation}\label{partition6}
Z(T,V,1)\simeq\left(\frac{240}{A^3\Tilde{\beta}^6}+\frac{1}{2}\right),
\end{equation}
or for a $N$-particles system, as
\begin{equation}\label{partition7}
Z(T,V,N)\simeq\left(\frac{240}{A^3m_0^6\beta^6}+\frac{1}{2}\right)^N,
\end{equation}
where we use the fact that $\Tilde{\beta}=\beta m_0$, and the factor $\frac{1}{2}$ was included for convenience (so we can plot the graph $C_V$ vs. $T$). 

Now, let's concentrate on the main thermodynamic properties, which are the Helmholtz free energy, the entropy, the mean energy, and the heat capacity, and are written in the following form \cite{Pacheco2,Oliveira4,Greiner,Pathria}
\begin{equation}\label{Helmholtz1}
F(T,V,N)=-\frac{1}{\beta}\ \mathsf{ln}[Z(T,V,N)],
\end{equation}
\begin{equation}\label{entropy1}
S(T,V,N)=k_B\beta^2\frac{\partial}{\partial\beta}F(T,V,N),
\end{equation}
\begin{equation}\label{energy1}
U(T,V,N)=-\frac{\partial}{\partial\beta}\mathsf{ln}[Z(T,V,N)],
\end{equation}
\begin{equation}\label{capacity1}
C_V(T,V,N)=-k_B\beta^2\frac{\partial}{\partial\beta}U(T,V,N).
\end{equation}

In this way, using the partition function \eqref{partition7}, the properties above be rewritten as
\begin{equation}\label{Helmholtz2}
\bar{F}\simeq-k_B T\ \mathsf{ln}\left(\frac{240k_B^6 T^6}{A^3 m_0^6}+\frac{1}{2}\right),
\end{equation}
\begin{equation}\label{entropy2}
\bar{S}\simeq k_B\left[\mathsf{ln}\left(\frac{240k_B^6 T^6}{A^3 m_0^6}+\frac{1}{2}\right)+\frac{2.880}{\left(\frac{A^3 m_0^6}{k_B^6 T^6}+480\right)}\right],
\end{equation}
\begin{equation}\label{energy2}
\bar{U}\simeq\frac{2.880}{\left(\frac{A^3 m_0^6}{k_B^7 T^7}+\frac{480}{k_B T}\right)},
\end{equation}
\begin{equation}\label{capacity2}
\bar{C}_V\simeq 2.880 k_B\frac{\left(\frac{7A^3 m_0^6}{k_B^6 T^6}+480\right)}{\left(\frac{A^3 m_0^6}{k_B^6 T^6}+480\right)},
\end{equation}
where $\bar{F}=\frac{F}{N}$ (Helmholtz free energy per particle), $\bar{S}=\frac{S}{N}$ (entropy per particle), $\bar{U}=\frac{U}{N}$ (mean energy per particle), and $\bar{C}_V=\frac{C_V}{N}$ (heat capacity per particle), respectively. Besides, here the pressure is zero, i.e., $P=-\frac{\partial F}{\partial V}=0$.

\subsection{The nonrelativistic case\label{subsec2}}

Now, let's consider the thermodynamic properties for the nonrelativistic case, which is the case where most of the phenomena of condensed matter physics or solid-state physics occur. However, we must first introduce the nonrelativistic energy spectrum for the NCQHE with AMM (spinless or $s=+1$), given by the following expression \cite{Oliveira5}
\begin{equation}\label{spectrum3}
\varepsilon_{n,m_l}=E_m+\tau\lambda\omega_c\left[n+\frac{1}{2}+\frac{\vert m_l\vert+m_l}{2}\right]>0.
\end{equation}

Considering a finitely degenerate spectrum (for $m_l\geq 0$), we have
\begin{equation}\label{spectrum4}
\varepsilon_{k}=\Bar{C}+Ck.
\end{equation}
where $k=n+m_l\geq 0$, $\Bar{C}=E_m+\frac{\tau\lambda\omega_c}{2}$ and $C=\tau\lambda\omega_c$.

So, since the one-particle partition function is written as follows
\begin{equation}\label{partition7}
Z(T,V,1)=\sum_{k=0}^\infty\Omega(\varepsilon_k)e^{-\beta\varepsilon_k},
\end{equation}
where the degeneracy $\Omega(\varepsilon_k)$ is also given by $\Omega(\varepsilon_k)=(k+1)^2$ (the same of the relativistic case), implies that we can rewrite this partition function as 
\begin{equation}\label{partition8}
Z(T,V,1)=\sum_{k=0}^\infty (k+1)^2e^{-[\bar{\bar{\beta}}+\tilde{\tilde{\beta}}k]},\end{equation}
where $\bar{\bar{\beta}}=\beta\Bar{C}$, and $\tilde{\tilde{\beta}}=\beta C$.

Using again the \textit{Euler-MacLaurin} formula, the function \eqref{partition8} becomes
\begin{equation}\label{partition9}
Z(T,V,1)=\left[\frac{1}{2}+\frac{(2+2\tilde{\tilde{\beta}}+\tilde{\tilde{\beta}}^2)}{\tilde{\tilde{\beta}}^3}-\frac{(2-\tilde{\tilde{\beta}})}{12}-\frac{(6\tilde{\tilde{\beta}}-6\tilde{\tilde{\beta}}^2+\tilde{\tilde{\beta}}^3)}{720}\right]e^{-\bar{\bar{\beta}}}+O(\tilde{\tilde{\beta}}^4),
\end{equation}
and for high temperatures, takes the form
\begin{equation}\label{partition10}
Z(T,V,1)\simeq\left(\frac{2}{\tilde{\Tilde{\beta}}^3}+\frac{1}{2}\right),
\end{equation}
or for a $N$-particles system, as
\begin{equation}\label{partition11}
Z(T,V,N)\simeq\left(\frac{2}{C^3\beta^3}+\frac{1}{2}\right)^N.
\end{equation} 

Therefore, using the partition function above, the nonrelativistic thermodynamics properties be written as
\begin{equation}\label{Helmholtz3}
\bar{F}\simeq-k_B T\ \mathsf{ln}\left(\frac{2k_B^3 T^3}{C^3}+\frac{1}{2}\right),
\end{equation}
\begin{equation}\label{entropy3}
\bar{S}\simeq k_B\left[\mathsf{ln}\left(\frac{2k_B^3 T^3}{C^3}+\frac{1}{2}\right)+\frac{12}{\left(\frac{C^3}{k_B^3 T^3}+4\right)}\right],
\end{equation}
\begin{equation}\label{energy3}
\bar{U}\simeq\frac{12}{\left(\frac{C^3}{k_B^4 T^4}+\frac{4}{k_B T}\right)},
\end{equation}
\begin{equation}\label{capacity3}
\bar{C}_V\simeq 48k_B\frac{\left(\frac{C^3}{k_B^3 T^3}+1\right)}{\left(\frac{C^3}{k_B^3 T^3}+4\right)},
\end{equation}
where $\bar{F}=\frac{F}{N}$ (Helmholtz free energy per particle), $\bar{S}=\frac{S}{N}$ (entropy per particle), $\bar{U}=\frac{U}{N}$ (mean energy per particle), and $\bar{C}_V=\frac{C_V}{N}$ (heat capacity per particle), with $P=-\frac{\partial F}{\partial V}=0$.
\section{Results and discussions\label{sec3}}

Here, we discuss in detail the (numerical) results through 2D graphs, where such graphs show the behavior of thermodynamic properties as a function of temperature, both for the nonrelativistic and relativistic cases. For the sake of simplicity, we consider $k_B=m_0=e=1$ (``unit constants''), and for convenience, we separate the allowable interval for the magnetic field $B$, given by $1<B<4$ (with $\theta=\eta=1$) \cite{Oliveira5}, into two parts ($1<B\leq 2.5$ and $2.5\leq B<4$), or into two symmetric sub-intervals (in relation to the point $B=2.5$).

\begin{figure}[!h]
\centering
\includegraphics[width=16.4cm]{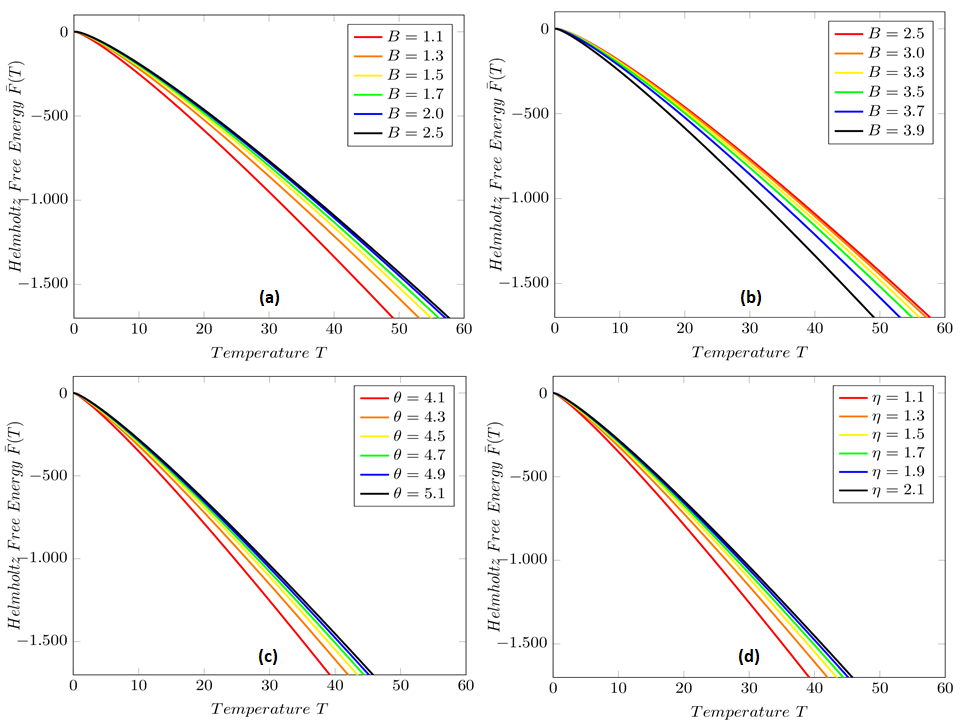}
\caption{Graph of $\bar{F}(T)$ vs. $T$ for six different values of $B$, $\theta$, and $\eta$ (relativistic case).}
\label{fig1}
\end{figure}

So, starting with the relativistic case, we have Fig. \ref{fig1}, which shows four graphs of Helmholtz free energy $\Bar{F}(T)$ as a function of temperature $T$ for six different values of $B$ (six for each sub-interval), $\theta$, and $\eta$. In general, all graphs of this Figure behave in a similar way, that is, $\Bar{F}(T)$ decreases logarithmically (or monotonically) as $T$ increases. However, each graph presents its own peculiarities. For example, in Fig. \ref{fig1}-(a) $\Bar{F}(T)$ increases as $B$ increases (for $1<B\leq 2.5$), while in Fig. \ref{fig1}-(b) $\Bar{F}(T)$ decreases as $B$ also increases (for $2.5\leq B<4$). In fact, this happens because the partition function does not depend on $E_m$ (or AMM), and therefore, presents a behavior analogous to the energy spectrum in the absence of $E_m$, where such spectrum increases from $B=1.1$ to $B=2.5$, and then decreases from $B=2.5$ to $B=3.9$ (the maximum value of the spectrum is in $B=2.5$)\cite{Oliveira5}. As a consequence of that, we have: $\Bar{F}(B=1.1)=\Bar{F}(B=3.9)$, $\Bar{F}(B=1.3)=\Bar{F}(B=3.7)$, $\Bar{F}(B=1.5)=\Bar{F}(B=3.5)$, $\Bar{F}(B=1.7)=\Bar{F}(B=3.3)$, and $\Bar{F}(B=2.0)=\Bar{F}(B=3.0)$.

Already in Figs. \ref{fig1}-(c) and \ref{fig1}-(d), we see that $\Bar{F}(T)$ increases as $\theta$ and $\eta$ increases (with $B=1$), i.e., the Helmholtz free energy increases as a function of noncommutativity. In addition, as the change of $\theta$ is equal to the change of $\eta$, i.e., $\Bar{F}(\Delta\theta)=\Bar{F}(\Delta\eta)$, or simply $\Bar{F}(\theta)=\Bar{F}(\eta)$, implies that the two graphs are exactly the same. Now, knowing that a change of $\bar{F}$ can be written as $\Delta \bar{F}\le -\bar{W}$, where $\bar{W}=\bar{W}_{non-PV}\ge 0$ is a type of work (per particle) done by the system (not mechanical work or non-PV work), implies that the smaller the value of $\Delta \bar{F}$, the larger the value of $\bar{W}$, and vice versa (the free energy of the system is depleted to do the work, i.e., is the energy available to do work) \cite{Greiner,Pathria}. On the other hand, since $\bar{F}$ can also be written as $\bar{F}=\bar{U}-T\bar{S}$ ($\bar{U}\le T\bar{S}$), implies that a $\bar{F}$ with increasingly negative values, more entropy is generated ($\bar{S}$ increases), and consequently, the system tends to equilibrium faster (faster heat exchange between the system and the thermal bath).

\begin{figure}[!h]
\centering
\includegraphics[width=16.4cm]{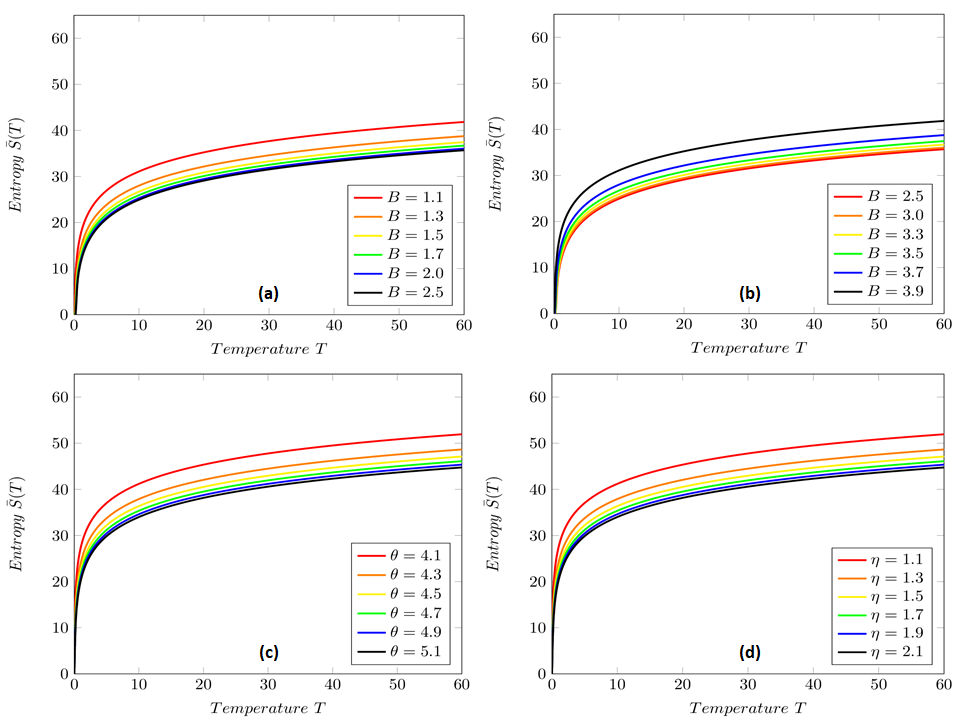}
\caption{Graph of $\bar{S}(T)$ vs. $T$ for six different values of $B$, $\theta$, and $\eta$ (relativistic case).}
\label{fig2}
\end{figure}

In Fig. \ref{fig2}, we have four graphs of entropy $\Bar{S}(T)$ as a function of temperature $T$ for six different values of $B$, $\theta$, and $\eta$. In general, all graphs of this Figure behave in a similar way, that is, $\Bar{S}(T)$ increases logarithmically (monotonically) as $T$ increases (as it should be, since $\Delta\Bar{S}\ge 0$). However, each graph presents its own peculiarities. For example, in Fig. \ref{fig2}-(a) $\Bar{S}(T)$ decreases as $B$ increases ($1<B\leq 2.5$), while in Fig. \ref{fig2}-(b) $\Bar{S}(T)$ increases as $B$ increases ($2.5\leq B<4$). As we mentioned earlier in the case of the Helmholtz free energy, this happens because the partition function does not depend on $E_m$. As a consequence of that, we have: $\Bar{S}(B=1.1)=\Bar{S}(B=3.9)$, $\Bar{S}(B=1.3)=\Bar{S}(B=3.7)$, $\Bar{S}(B=1.5)=\Bar{S}(B=3.5)$, $\Bar{S}(B=1.7)=\Bar{S}(B=3.3)$, and $\Bar{S}(B=2.0)=\Bar{S}(B=3.0)$. Already in Figs. \ref{fig2}-(c) and \ref{fig2}-(d), we see that $\Bar{S}(T)$ decreases as $\theta$ and $\eta$ increases (with $B=1$), i.e., the entropy decreases as a function of noncommutativity. In addition, as the change of $\theta$ is equal to the change of $\eta$, i.e., $\Bar{S}(\Delta\theta)=\Bar{S}(\Delta\eta)$, or simply $\Bar{S}(\theta)=\Bar{S}(\eta)$, implies that the two graphs are exactly the same.

\begin{figure}[!h]
\centering
\includegraphics[width=16.4cm]{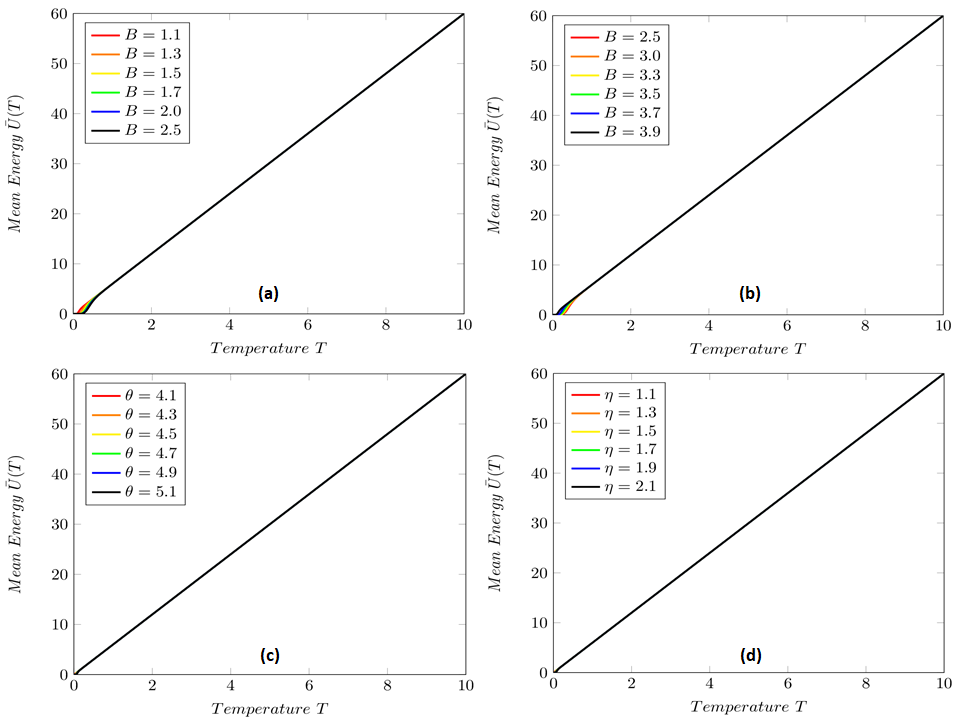}
\caption{Graph of $\bar{U}(T)$ vs. $T$ for six different values of $B$, $\theta$, and $\eta$ (relativistic case).}
\label{fig3}
\end{figure}

In Fig. \ref{fig3}, we have four graphs of mean energy $\Bar{U}(T)$ as a function of temperature $T$ for six different values of $B$, $\theta$, and $\eta$. In general, all graphs of this Figure behave in a similar way, that is, $\Bar{U}(T)$ increases linearly as $T$ increases, with $\bar{U}(T)=\bar{F}+T\bar{S}=6k_BT$ (restoring the constant $k_B$). In particular, according to the equipartition theorem, which states that each degree of freedom contributes with $\frac{1}{2}k_B T$ per particle to the mean energy (total mean internal energy) \cite{Greiner,Pathria}, this result implies that the system has twelve degrees of freedom (twelve independent ways to store energy) since $U(T)=\frac{f}{2}Nk_B T$, where $f$ represents the number of degrees of freedom. However, unlike the two previous thermodynamic properties, here the mean energy only depends on the temperature (as it should be), and therefore, the four graphs are exactly the same (for high temperatures).

\begin{figure}[!h]
\centering
\includegraphics[width=16.4cm]{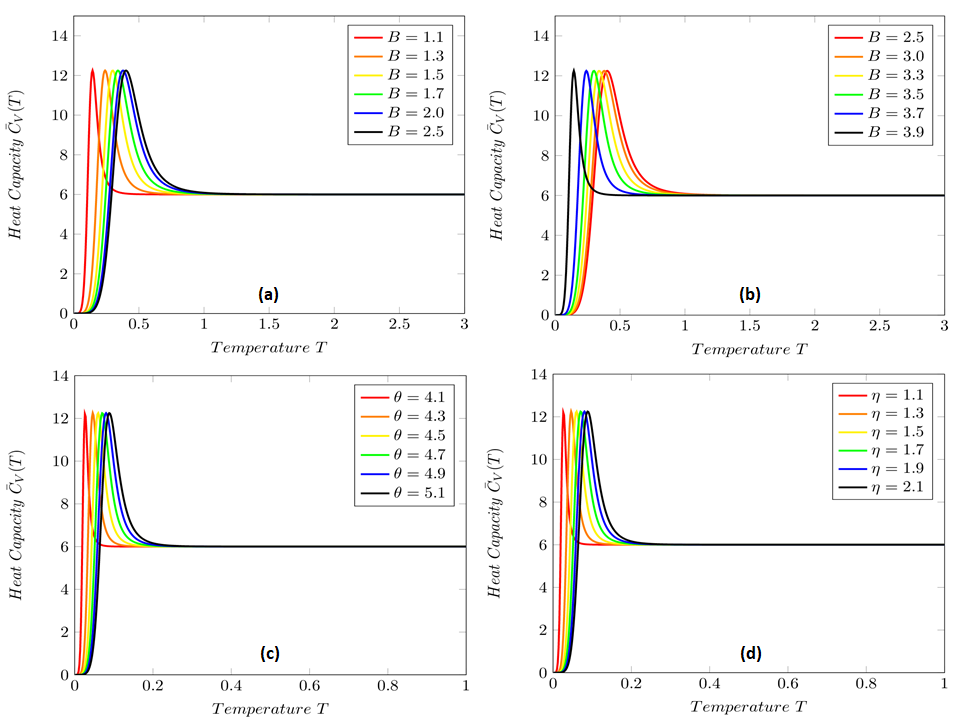}
\caption{Graph of $\bar{C}_V(T)$ vs. $T$ for six different values of $B$, $\theta$, and $\eta$ (relativistic case).}
\label{fig4}
\end{figure}

Already in Fig. \ref{fig4}, we have four graphs of heat capacity $\Bar{C}_V(T)$ as a function of temperature $T$ for six different values of $B$, $\theta$, and $\eta$. In general, all graphs of this Figure behave in a similar way, that is, $\Bar{C}_V(T)$ tends to a constant as $T$ increases, given by $\Bar{C}_V=6 k_B$ (restoring the constant $k_B$). In particular, this result satisfies the so-called
\textit{Dulong-Petit} law, where it states that the heat capacity is a constant for high temperatures \cite{Greiner,Pathria}. However, comparing the four graphs in relation to the three physical parameters ($B$, $\theta$, and $\eta$), we see that the heat capacity reaches a constant value much faster in the last two graphs (which are exactly the same). Furthermore, with respect to the maximum value (maximum peak) of $\bar{C}_V$ \cite{Greiner,Pathria}, we have $\bar{C}_V\approx 12 k_B$ (i.e., a little more than double the value for large $T$), where this value shifts to larger values of $T$ as the field $B$ goes from $1.1$ to $2.5$ (or $\theta$ and $\eta$ with values increasing), or shifts to smaller values of $T$ as the field $B$ goes from $2.5$ to $3.9$.

\begin{figure}[!h]
\centering
\includegraphics[width=16.4cm]{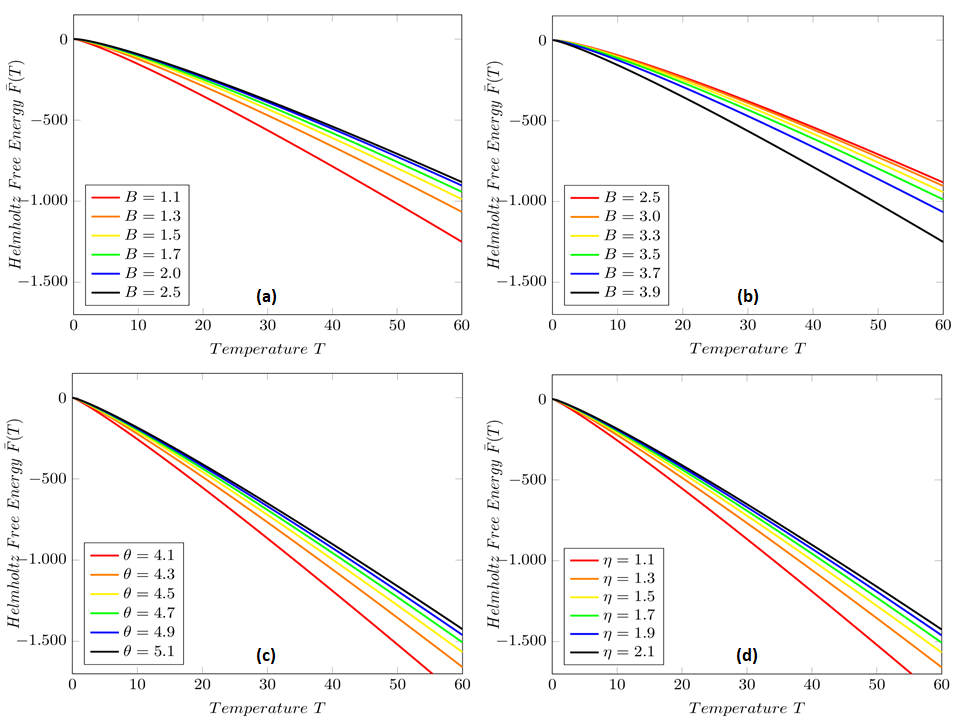}
\caption{Graph of $\bar{F}(T)$ vs. $T$ for six different values of $B$, $\theta$, and $\eta$ (nonrelativistic case).}
\label{fig5}
\end{figure}

Now, for the nonrelativistic case, we have Fig. \ref{fig5}, which shows four graphs of Helmholtz free energy $\Bar{F}(T)$ as a function of temperature $T$ for six different values of $B$, $\theta$, and $\eta$. So, analogous to the relativistic case, $\Bar{F}(T)$ also decreases logarithmically as $T$ increases, increases as $B$ increases from $B=1.1$ to $B=2.5$ (Fig. \ref{fig5}-(a)), or as $\theta$ and $\eta$ increase (Figs. \ref{fig5}-(c) and \ref{fig5}-(d)), decreases as $B$ increases from $B=2.5$ to $B=3.9$ (Fig. \ref{fig5}-(b)), where $\Bar{F}(B=1.1)=\Bar{F}(B=3.9)$, $\Bar{F}(B=1.3)=\Bar{F}(B=3.7)$, $\ldots$, and $\Bar{F}(\theta)=\Bar{F}(\eta)$. However, comparing both relativistic and nonrelativistic cases, with $\Delta\Bar{F}\le -\Bar{W}$, we note that $\Bar{F}$ is much smaller for the relativistic case ($\Bar{F}^{rel}\ll\Bar{F}^{nonrel}$), consequently, $\Bar{W}$ is much larger for the relativistic case ($\Bar{W}^{rel}\gg\Bar{W}^{nonrel}$).

\begin{figure}[!h]
\centering
\includegraphics[width=16.4cm]{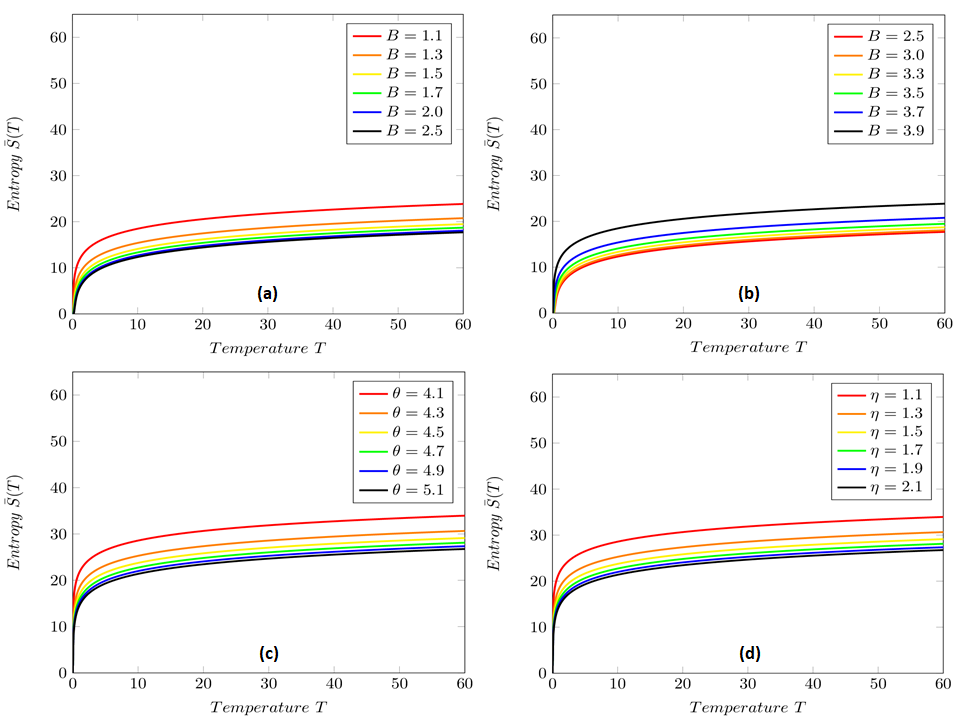}
\caption{Graph of $\bar{S}(T)$ vs. $T$ for six different values of $B$, $\theta$, and $\eta$ (nonrelativistic case).}
\label{fig6}
\end{figure}

In Fig. \ref{fig6}, we have four graphs of entropy $\Bar{S}(T)$ as a function of temperature $T$ for six different values of $B$, $\theta$, and $\eta$. So, analogous to the relativistic case, $\Bar{S}(T)$ also increases logarithmically as $T$ increases, decreases as $B$ increases from $B=1.1$ to $B=2.5$ (Fig. \ref{fig6}-(a)), or as $\theta$ and $\eta$ increase (Figs. \ref{fig6}-(c) and \ref{fig6}-(d)), increases as $B$ increases from $B=2.5$ to $B=3.9$ (Fig. \ref{fig6}-(b)), where $\Bar{S}(B=1.1)=\Bar{S}(B=3.9)$, $\Bar{S}(B=1.3)=\Bar{S}(B=3.7)$, $\ldots$, and $\Bar{S}(\theta)=\Bar{S}(\eta)$. However, comparing both relativistic and nonrelativistic cases, with $\Delta\Bar{S}\ge\Bar{Q}/T$, we note that the entropy is larger for the relativistic case ($\Bar{S}^{rel}>\Bar{S}^{nonrel}$), consequently, the quantity of heat (per particle) absorbed (from the thermal bath) by the system is larger for the relativistic case ($\Bar{Q}^{rel}>\Bar{Q}^{nonrel}$).

\begin{figure}[!h]
\centering
\includegraphics[width=16.4cm]{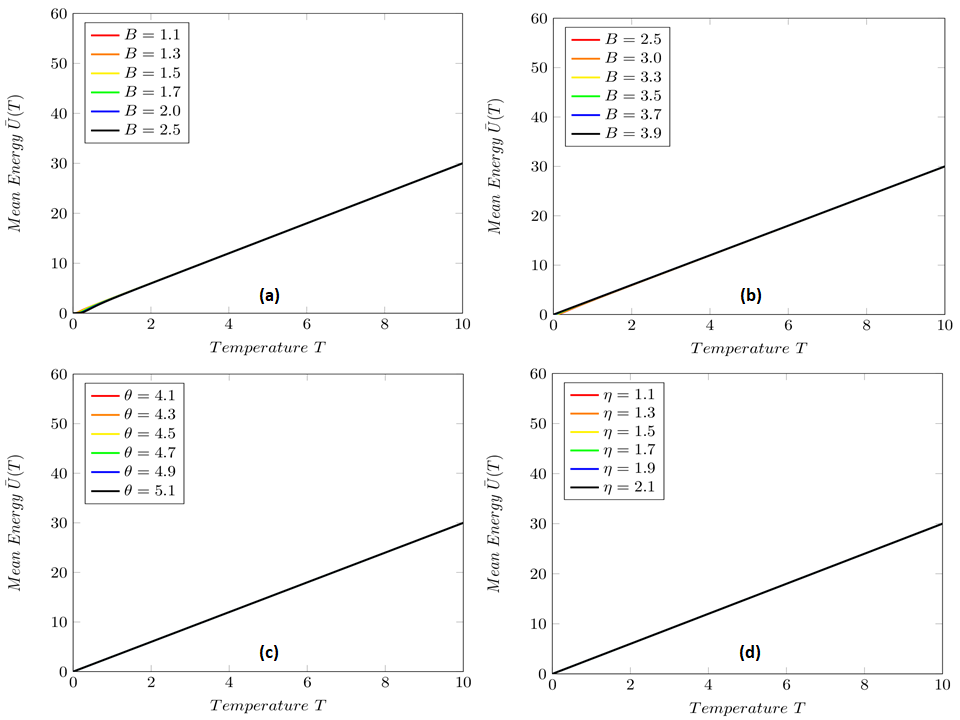}
\caption{Graph of $\bar{U}(T)$ vs. $T$ for six different values of $B$, $\theta$, and $\eta$ (nonrelativistic case).}
\label{fig7}
\end{figure}

In Fig. \ref{fig7}, we have four graphs of mean energy $\Bar{U}(T)$ as a function of temperature $T$ for six different values of $B$, $\theta$, and $\eta$. So, analogous to the relativistic case, $\Bar{U}(T)$ also increases linearly as $T$ increases, and only depends on the temperature, i.e., $\Bar{U}(T)=3k_B T$. In particular, this result is equal to the value of the mean energy of the Debye solid, or of an ultrarelativistic ideal gas \cite{Greiner,Pathria}. Besides, we note that $\Bar{U}(T)$ for the relativistic case is twice of the nonrelativistic case ($\Bar{U}^{rel}=2\Bar{U}^{nonrel}$).

\begin{figure}[!h]
\centering
\includegraphics[width=16.4cm]{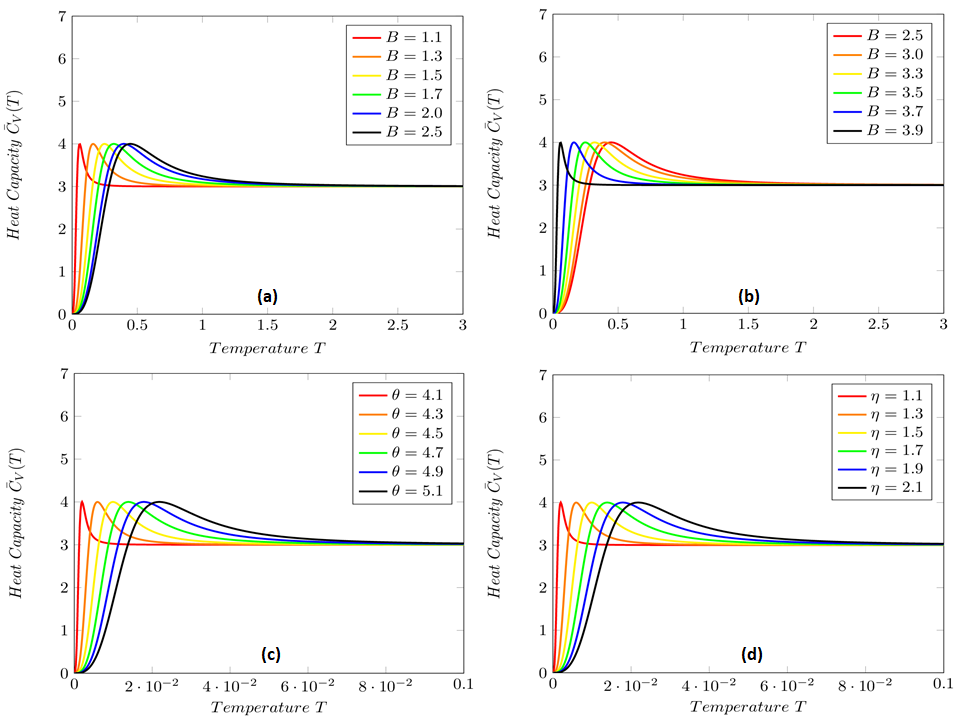}
\caption{Graph of $\bar{C}_V(T)$ vs. $T$ for six different values of $B$, $\theta$, and $\eta$ (nonrelativistic case).}
\label{fig8}
\end{figure}

Already in Fig. \ref{fig8}, we have four graphs of heat capacity $\Bar{C}_V(T)$ as a function of temperature $T$ for six different values of $B$, $\theta$, and $\eta$. So, analogous to the relativistic case, $\Bar{C}_V(T)$ also tends to a constant as $T$ increases, given by $\Bar{C}_V=3k_B$. In particular, this result is equal to the value of the heat capacity of the Debye solid, or of an ultrarelativistic ideal gas \cite{Greiner,Pathria}. Besides, we note that $\Bar{C}_V(T)$ for the relativistic case is twice of the nonrelativistic case ($\Bar{C}_V^{rel}=2\Bar{C}_V^{nonrel}$). In fact, this occurs because we have $\Bar{U}^{rel}=2\Bar{U}^{nonrel}$. Furthermore, with respect to the maximum value (maximum peak) of $\bar{C}_V$, we have $\bar{C}_V=4 k_B$ (here it is not twice the $\bar{C}_V$ for large $T$ as in the relativistic case), where this value shifts to larger values of $T$ as the field $B$ goes from $1.1$ to $2.5$ (or $\theta$ and $\eta$ with values increasing), or shifts to smaller values of $T$ as the field $B$ goes from $2.5$ to $3.9$.

Last but not least, it is worth mentioning that since the partition function in both nonrelativistic and relativistic cases does not depend on $E_m$ (or AMM), implies that the thermodynamic properties also do not depend. In fact, there is no influence of $E_m$ (or AMM) on the thermodynamic properties because such physical quantity is not ``linked'' to the quantum number $k$, unlike $B$ (or $\omega_c$), $\theta$, and $\eta$.

\section{Conclusion\label{conclusion}}

In this paper, we study the thermodynamic properties of the NCQHE with AMM for both relativistic and nonrelativistic cases in the high temperatures regime. To achieve this, we use the canonical ensemble for a set of noninteracting $N$-particles in contact with a thermal bath at a constant temperature $T$. Next, we explicitly determine the thermodynamic properties of our interest, namely: the Helmholtz free energy $\bar{F}$, the entropy $\bar{S}$, the mean energy $\bar{U}$, and the heat capacity $\bar{C}_V$. Also, in order to perform the calculations of these properties, we work with the \textit{Euler-MacLaurin} formula to construct the partition function of the system. Then, to better analyze and discuss in detail the results, we plotted the graphs of the behavior of thermodynamic properties as a function of temperature for six different values of $B$, $\theta$, and $\eta$ (which are the three relevant physical parameters). 

So, in general we note that $\Bar{F}$ decreases logarithmically as $T$ increases, increases as $B$ increases from $B=1.1$ to $B=2.5$, or as $\theta$ and $\eta$ increase, and decreases as $B$ increases from $B=2.5$ to $B=3.9$. However, comparing both relativistic and nonrelativistic cases, we note that $\Bar{F}$ is much smaller for the relativistic case ($\Bar{F}^{rel}\ll\Bar{F}^{nonrel}$), consequently, the work $\Bar{W}$ (work done by the system) is much larger for the relativistic case ($\Bar{W}^{rel}\gg\Bar{W}^{nonrel}$). Already for the entropy, in general, we note that $\Bar{S}(T)$ increases logarithmically as $T$ increases, decreases as $B$ increases from $B=1.1$ to $B=2.5$, or as $\theta$ and $\eta$ increase, and increases as $B$ increases from $B=2.5$ to $B=3.9$. However, comparing both relativistic and nonrelativistic cases, we note that the entropy is larger for the relativistic case ($\Bar{S}^{rel}>\Bar{S}^{nonrel}$), consequently, the quantity of heat $\Bar{Q}$ absorbed by the system is larger for the relativistic case ($\Bar{Q}^{rel}>\Bar{Q}^{nonrel}$).

Now, with respect to the mean energy, in general, $\Bar{U}(T)$ increases linearly as $T$ increases, and only depends on the temperature, where $\Bar{U}^{rel}=6k_B T$, and $\Bar{U}^{nonrel}=3k_B T$, i.e., $\Bar{U}$ for the relativistic case is twice of the nonrelativistic case. In particular, $\Bar{U}^{nonrel}$ is equal to the value of the mean energy of the Debye solid (or an ultrarelativistic ideal gas). Already for the heat capacity, in general, we note that $\Bar{C}_V(T)$ tends to a constant as $T$ increases, where $\Bar{C}_V^{rel}=6k_B$, and $\Bar{C}_V^{nonrel}=3k_B$ (the relativistic case is twice of the nonrelativistic case), and therefore, satisfies the \textit{Dulong-Petit} law. Besides, $\Bar{C}_V^{nonrel}$ is equal to the value of the mean energy of the heat capacity of the Debye solid (or an ultrarelativistic ideal gas). Finally, we verify that there is no influence of $E_m$ (or AMM) on the thermodynamic properties of the system, i.e., the partition function does not depend on $E_m$. In fact, this occurs because such a physical quantity is not ``linked'' to the quantum number $k$, unlike $B$ (or $\omega_c$), $\theta$, and $\eta$.

\section*{Acknowledgments}

\hspace{0.5cm}The authors would like to thank the Conselho Nacional de Desenvolvimento Cient\'{\i}fico e Tecnol\'{o}gico (CNPq) for financial support.

\end{document}